\documentclass[showpacs,groupedaddress,assymb,amsmath]{revtex4}
\usepackage{}
\usepackage{graphicx}
\usepackage{amssymb}
\usepackage{amsmath}

\begin{document}
\title{Quantum optical diode with  semiconductor microcavities}
\author{H. Z. Shen$^1$, Y. H. Zhou$^1$, and
X. X. Yi$^2$\footnote{Corresponding  address: yixx@nenu.edu.cn}}
\affiliation{$^1$ School of Physics and Optoelectronic Technology\\
Dalian University of Technology, Dalian 116024 China\\
$^2$ Center for Quantum Sciences and School of Physics, Northeast
Normal University, Changchun 130024, China }

\date{\today}

\begin{abstract}
The semiconductor diode, which acts as an electrical rectifier and
allows unidirectional electronic transports, is the key to
information processing in integrated circuits. Analogously, an
optical rectifier (or diode)  working  at specific target
wavelengths has recently becomes a dreaming  device in optical
communication and signal processing. In this paper, we propose a
scheme to realize   an optical diode   for photonic transport at the
level of few photons. The system consists of two spatially
overlapping single-mode semiconductor microcavities coupled via
${\chi ^{(2)}}$ nonlinearities. The photon blockade is predicted to
take place in this system. These photon blockade effects can be
achieved by tuning the frequency of the input laser field (driving
field). Based on those blockades, we derive analytically the single-
and two-photon current in terms of zero and finite-time delayed
two-order correlation function.  The results suggest that the system
can serve as an single- and two-photon quantum optical diodes which
allow transmission of photons in one direction much more efficiently
than in the other.
\end{abstract}

 \maketitle
\section{Introduction}
The electrical diode is a two-terminal electronic device  with
asymmetric conductance, it has low   resistance to current flow in
one direction, and high   resistance in  the opposite direction. The
study on electrical diodes can be traced  back to more than one
century ago, when the first device  enables the rectification of
current flux. Motivated by the significant rectifying capabilities
of electric diodes, considerable efforts have been made  to
investigations of the rectification of other energy forms, for
example, thermal
flux \cite{Li931843012004,Li951043022005,Chang31411212006,Kobayashi951719052009,
Terraneo88094302,Scheibner100830162008} and solitary
waves \cite{Nesterenko95158702}.

Efforts have also been made for the optical diodes. Optical diodes,
also known as optical isolators \cite{Aplet35441964}, are an optical
rectifier working  at specific target wavelengths, they allow
propagation of photon signal in one direction and block that in the
opposite direction. Motivated by potential applications in quantum
network of light, various possible solid-state optical diodes have
been proposed, including the diodes from  standard bulk Faraday
rotators \cite{Yeh1988,Saleh2007}, the diodes  integrated on a chip
\cite{Kang55492011,Kamal73112011,Fan3354472011,Bi57582011,Yu3912009},
and diodes operating in optically controllable way
\cite{Kang55492011,Alberucci3316412008,Biancalana1040931132008}, as
well as the diodes at low field intensities or even at the
single-photon levels \cite{Shen1071739022011}.

The physics behind the optical diode is the  breaking of time
reversal symmetry, which  is typically achieved via  acoustic
rectifiers \cite{Liang1031043012009,Liang99892010}, moving photonic
crystal \cite{Wang1100939012013}, spin-photon entangling
\cite{Flindt982405012007} and  few-photon tunneling
\cite{Roy811551172010,Nikoghosyan1031636032009,Blakesley940674012005}.
Thanks to the classical level these schemes work at, they have now
been attained in different configurations on-chip
\cite{Bi57582011,Fan3354472012,Lira1090339012012}.

With the recent advances in quantum photonic technologies at the
single-photon levels \cite{O'Brien36872009}, researchers make a step
further in the study of  optical diode--{\it quantum optical diode},
tunable single- or two-photon quantum rectification is likely to
play an important role in this case analogous to the classical
electrical diodes in current microchips. A quantum optical rectifier
is a two terminal, spatially nonreciprocal device that allows
unidirectional propagation of single- or few quanta of
(electromagnetic) energy at a fixed signal frequency and amplitude
\cite{Mascarenhas130764932013}.  Up to now, only few proposals have
been proposed
 \cite{Shen1071739022011,Mascarenhas130764932013} in fully quantum
regime.

In this paper, we propose  a fully quantum diode with  two coupled
semiconductor microcavities driven by external fields. The merit of
this proposal is to unify the unidirectional transport of photon and
photon conversion. The system consists of two spatially overlapping
single-mode cavities with frequencies ${\omega _b}$ and ${\omega
_a}$, respectively. The two cavities  are coupled by   ${\chi
^{(2)}}$ nonlinearities that mediate the conversion of a photon in
cavity $b$ to two photons in cavity $a$ and vice
versa\cite{Irvine960574052006,Agarwal73522,Carusotto6049071999,
Majumdar872353192013,Liu1110836012013}. The physics behind this
proposal is the photon blockade. When  cavity $a$ is pumped, single
photon blockade prevails in the system,  the single photon who
occupies the Fock state first  blockades the generation of more
photons. When the pumping is on cavity $b$, due to the existence of
the two-photon eigenstate of the system, two-photon blockade
dominates, blockading more photons in cavity $b$.

The single-photon blockade was already observed in the standard
cavity quantum electrodynamics (QED)
\cite{Tian46R68011992,Werner610118011999,Brecha5923921999,
Rebic14901999,Rebic650638042002,
Kim3975001999,Smolyaninov881874022002,Bamba83021802(R)2011,
Liew1041836012010,Miranowicz870238092013}, optomechanical systems
\cite{Rabl1070636012011,Liao820538362010,Komar870138392013,Liao880238532013}
and in circuit QED system
\cite{Hoffman1070536022011,Lang1062436042011}. The proposal
presented here  is to use these techniques and show the single- and
two-photon blockade effect by changing  the intensity and frequency
of the driving field.

The remainder of the paper is organized as follows. In Sec. {\rm
II}, we introduce the system and present a model to describe the
system. The single- and two-photon blockade is predicted and
discussed.  In Sec. {\rm III}, in terms of the two-order correlation
function and photon number statistics, we analyze  a specific
single-photon and two-photon diode. In Sec. {\rm IV}, we investigate
the rectification  of the diode via  two-time correlation functions.
Discussion and conclusions are given in Sec. {\rm V}.
\section{Model and photon blockade}
Throughout this work, we adopt the International System (SI) of
units \cite{Boyd2008}. The nonlinear optical response of dielectric
material to an electric field   is given by
\begin{equation}
\begin{aligned}
{D_i}(\mathbf{r},t) =& \sum\limits_{jkm} {\{ {\varepsilon _0}
{\varepsilon _{ij}}(\mathbf{r}){E_j}(\mathbf{r},t) +
{\varepsilon _0}[\chi _{ijk}^{(2)}(\mathbf{r}){E_j}(\mathbf{r},t)} \\
& \times {E_k}(\mathbf{r},t) + \chi _{ijkm}^{(3)}(\mathbf{r}){E_j}
(\mathbf{r},t){E_k}(\mathbf{r},t)\\
& \times {E_m}(\mathbf{r},t) +  \cdot  \cdot  \cdot ]\},
 \label{Di}
\end{aligned}
\end{equation}
which  defines the relative dielectric permettivity  tensor of the
medium, ${\varepsilon _{ij}}(\mathbf{r}) = {\delta _{ij}} + \chi
_{ij}^{(1)}(\mathbf{r})$. We will consider only the nonlinear
response up to second order in the electromagnetic field, i.e., we
assume $\chi _{ijkm}^{(3)}(\mathbf{r})=0$ and only take the optical
nonlinear effects caused by $\chi^{(2)}_{ijk}$  in Eq.~(\ref{Di}) into
account. We assume the material is isotropic, i.e., ${\varepsilon
_{ij}}(\mathbf{r}) \to \varepsilon (\mathbf{r})$ is a spatially
dependent scalar quantity. The canonical quantization can be done by
expressing the   field operators  as
\begin{equation}
\begin{aligned}
\hat E(\mathbf{r},t) = \sum\limits_{j = a,b} {\sqrt {\frac{{\hbar
{\omega _j}}}{{2{\epsilon _0}}}} }  [{{\hat h}_j}\frac{{{{\vec \phi
}_j}(\mathbf{r})}}{{\sqrt {\varepsilon (\mathbf{r})} }}{e^{ -
i{\omega _j}t}} + H.c.], \label{Eab}
\end{aligned}
\end{equation}
and $\hat B({\bf{r}},t) =  - \nabla  \times \hat
E({\bf{r}},t)/{\omega _0},$ with H.c. standing  for the Hermitian
conjugate. Here ${{\hat h}_a} = \hat a$ and ${{\hat h}_b} = \hat b$
denote the photon destruction operators   in the two cavity with
frequencies ${\omega _a}$ and ${\omega _b} = 2{\omega _a}$,
respectively.  For each  cavity mode, the three-dimensional cavity
field $\phi_j({\bf{r}})$ must be normalized by $\int {{{\left|
{{{\vec \phi }_j}({\bf{r}})} \right|}^2}{d^3}{\bf{r}} = 1}$ (j=a,b).
By the energy density formula in classical electrodynamics, ${{\rm
H}_{em}} = \frac{1}{2}\int {[\mathbf{E}(\mathbf{r},t) \cdot }
\mathbf{D}(\mathbf{r},t) + \int { \mathbf{H}(\mathbf{r},t) \cdot }
\mathbf{B}(\mathbf{r},t)]{d^3}\mathbf{r}$, where $
\mathbf{H}(\mathbf{r},t) = \mathbf{B}(\mathbf{r},t)/{\mu _0}$, an
interaction  Hamiltonian  in the second quantization can be obtained
 \cite{Irvine960574052006,Carusotto6049071999}
\begin{eqnarray}
\hat H_0 = \hbar {\omega _a}{{\hat a}^\dag }\hat a +  \hbar {\omega
_b}{{\hat b}^\dag }\hat b + \hbar \Omega ({{\hat b}^\dag }{{\hat
a}^2} + H.c.) \label{H}
\end{eqnarray}
where the nonlinear interaction coefficient is defined by,
\begin{eqnarray}
\Omega  = \sqrt {\frac{{{\hbar ^3}\omega _a^2{\omega _b}}}{{8{\varepsilon _0}}}} \sum\limits_{ijk} {\int {\frac{{\chi _{ijk}^{(2)}({\bf{r}})}}{{{{[\varepsilon ({\bf{r}})]}^{1.5}}}}\phi _{i,a}^ * ({\bf{r}})\phi _{j,a}^ * ({\bf{r}}){\phi _{k,b}}({\bf{r}}){d^3}{\bf{r}}} } .
\label{omega}
\end{eqnarray}

\begin{figure}[h]
\centering
\includegraphics[angle=0,width=0.499\textwidth]{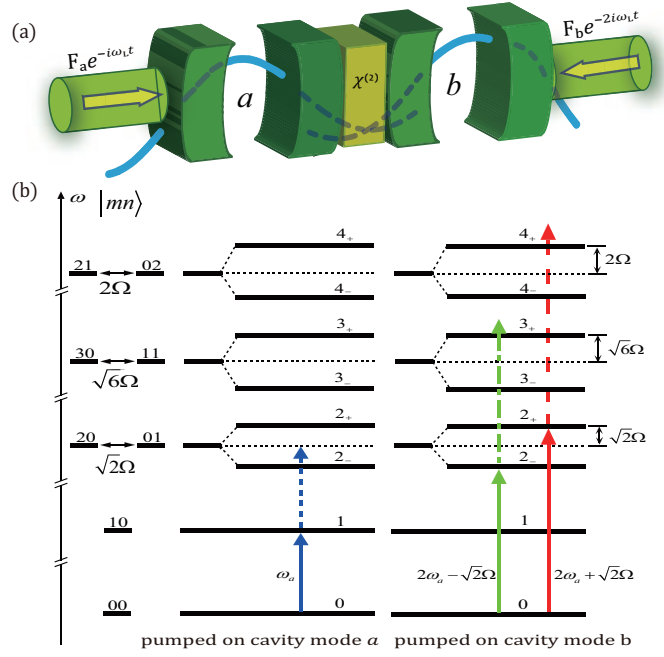}
\caption{(Color online) (a) Illustration of the setup. The two
cavities driven by external fields(pumping)  are  coupled by
$\chi^{2}$ nonlinearities. The pumping on the cavity $a$ is resonant
with  the field inside  cavity $a$, while the pumping on cavity $b$
is at $2\omega_a+\sqrt 2\Omega$ or $2\omega_a-\sqrt 2\Omega$. (b)
Level diagram for the two coupled cavities with ${\omega _b} =
2{\omega _a}$. States are labeled by $\left| {mn} \right\rangle$
with the first(second) number denoting the photon number   in cavity
$a$ (cavity $b$). The coupling $\Omega$ splits the degeneracy of
states $\left| {mn} \right\rangle$ and $\left| {m-2n+1}
\right\rangle$ or $\left| {m+2n-1} \right\rangle$. The arrows show
the frequency  of the driven field.} \label{setupsssss:}
\end{figure}
For the scheme to work, external driving for cavity $a$ or $b$ is
essential. We illustrate the setup in Fig.~\ref{setupsssss:} (a), the
driving  frequency is denoted  by $\omega_L$ for cavity $a$ and
$2\omega_L$ for cavity $b$, respectively.  $F$ stands for the
driving strength. The corresponding Hamiltonian is,
\begin{eqnarray}
{{\hat H}_{dr}}(t) = {{\hat H}_0} + \hbar F{{\hat h}_j}{e^{ -
i{\lambda _j}{\omega _L}t}} + \hbar {F^*}
\hat h_j^\dag {e^{i{\lambda _j}{\omega _L}t}},\label{hj}
\end{eqnarray}
where ${\lambda _a} = 1$  and ${\lambda _b} = 2$. In a rotating
frame defined by $\hat U_S(t) = \exp [i{\omega _L}t({{\hat a}^\dag
}\hat a + 2{{\hat b}^\dag }\hat b)]$, the Hamiltonian takes,
\begin{equation}
\begin{aligned}
{{\hat H}_S} =& \hbar {\Delta _a}{{\hat a}^\dag }\hat a + \hbar {\Delta _b}{{\hat b}^\dag }\hat b + \hbar \Omega ({{\hat b}^\dag }{{\hat a}^2} + {{\hat a}^{\dag 2}}\hat b) + \hbar F{{\hat h}_j}\\
& + \hbar {F^*}\hat h_j^\dag .
\label{Hj}
\end{aligned}
\end{equation}
${\Delta _a} = {\omega _a} - {\rm{ }} {\omega _L}$ and ${\Delta _b}
= {\omega _b} - 2{\omega _L}$ define the  detunings of cavity $a$
and  cavity $b$ modes from the driving laser, respectively.

The nonlinear terms proportional to $\Omega$ in Eq.~(\ref{Hj})
describe coherent photon exchange between the two optical cavity
modes.  The resulting low-energy level diagram is shown in
Fig.~\ref{setupsssss:} (b), where $\left| {{\rm{mn}}} \right\rangle
$ as before represents a state with $m$ and $n$ photons in the
cavity $a$ and $b$, respectively. In the absence of driving  fields,
we diagonalize ${{\hat H_S}}$  with ${\omega _b} = 2{\omega _a}$ and
$\omega_L=\omega_a$. The ground state and low excited state are
listed below,
\begin{equation}
\begin{aligned}\left| 0 \right\rangle  =& \left| {00} \right\rangle ,\\
\left| 1 \right\rangle  =& \left| {10} \right\rangle ,\\
\left| {{2_ \pm }} \right\rangle  =&
\frac{1}{{\sqrt 2 }}\left| {01} \right\rangle  \pm \left| {20} \right\rangle ,\\
\left| {{3_ \pm }} \right\rangle  =&
\frac{1}{{\sqrt 2 }}\left| {11} \right\rangle  \pm \left| {30} \right\rangle ,\\
\left| {{4_ \pm }} \right\rangle  =&
\frac{1}{{\sqrt 2 }}\left| {02} \right\rangle  \pm \left| {21} \right\rangle .
\label{01}
\end{aligned}
\end{equation}
The driving field couples all  states which differ from each other
by a single photon. In the following sections, we will restrict
ourself to consider a very weak driving filed such that the Hilbert
space can be truncated to low energy levels, as listed in Eq.
(\ref{01}).

In addition to the coherent evolution governed by the Hamiltonian
${{\hat H}_S}$, we introduce  cavity losses   to the system, and the
dynamics of system is described by,
\begin{eqnarray}
\dot \rho  =  - i[{{\hat H}_S},\rho ]
+ {\kappa _a}{\cal D}(\hat a)\rho  + \kappa {\cal D}(\hat b)\rho ,\label{rho}
\end{eqnarray}
where $\hat H_S$ is given by Eq.~(\ref{Hj}), $\kappa$  and
$\kappa_a$ denotes loss  rates for the two cavities, respectively,
the superoperator $\mathcal{D}(\hat o)=\hat o\rho {{\hat o}^\dag } -
\frac{1}{2}{{\hat o}^\dag }\hat o\rho  - \frac{1}{2}\rho {{\hat
o}^\dag }\hat o$. According to the Markovian input-output formalism
of Collett and Cardiner
\cite{Collett3013861984,Cardiner3137611985,Shen880338352013}, the
input, output, and intracavity fields are linked by the input-output
relation,
\begin{equation}
\begin{aligned}
{{\hat a}_{out}}(t) =& {{\hat a}_{in}}(t) + \sqrt {\kappa_a} \hat a(t),\\
{{\hat b}_{out}}(t) =& {{\hat b}_{in}}(t) + \sqrt \kappa \hat b(t).
\label{inout}
\end{aligned}
\end{equation}
For an arbitrary state of the input modes, correlations of the
output fields would depend on cross correlations between the input
and intracavity fields, which in turn would require to model the
input fields together with the system dynamics. However, if we
assume that only classical driving fields is added to the quantum
vacuum of the input-output channels, then all normally ordered cross
correlations between intracavity and input modes vanish, and
correlations in the output channels can be expressed as functions of
intracavity correlations only. With this assumption, the average
output current  (or photon stream) at time $t$ can be formally given
by,
\begin{eqnarray}
{N_j}(t) = {\kappa _j}Tr[\hat h_j^\dag {{\hat h}_j}\rho (t)].\label{Nj}
\end{eqnarray}
We will set ${\kappa _b} \equiv \kappa $ hereafter.

Now we discuss photon-blockade effect.  This effect can be
characterized by the second order correlation function with no
time-delay \cite{Verger731933062006,London2003}
\begin{figure}[t]
\centerline{
\includegraphics[width=8.9cm]{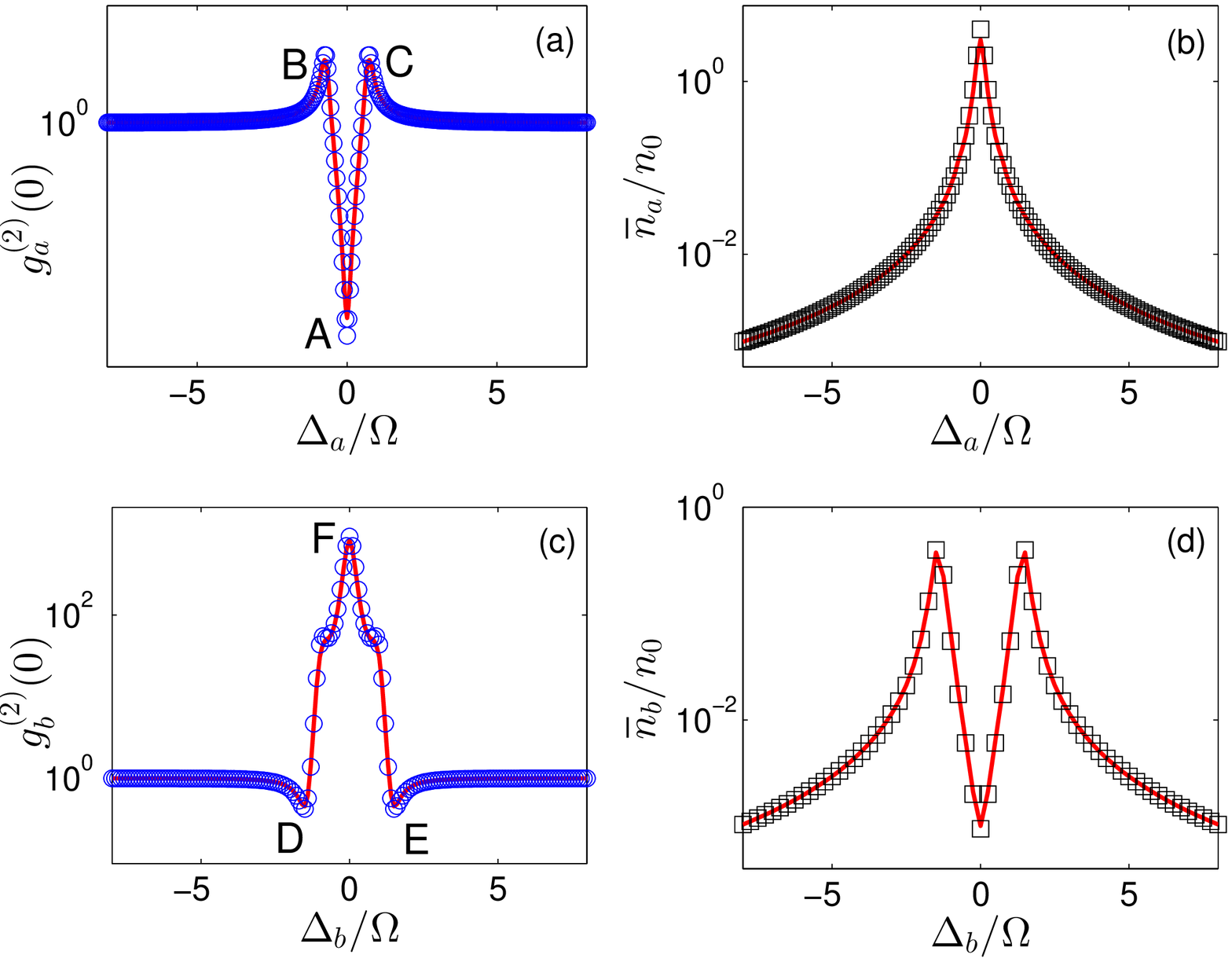}}
\centerline{
\includegraphics[width=7.5cm]{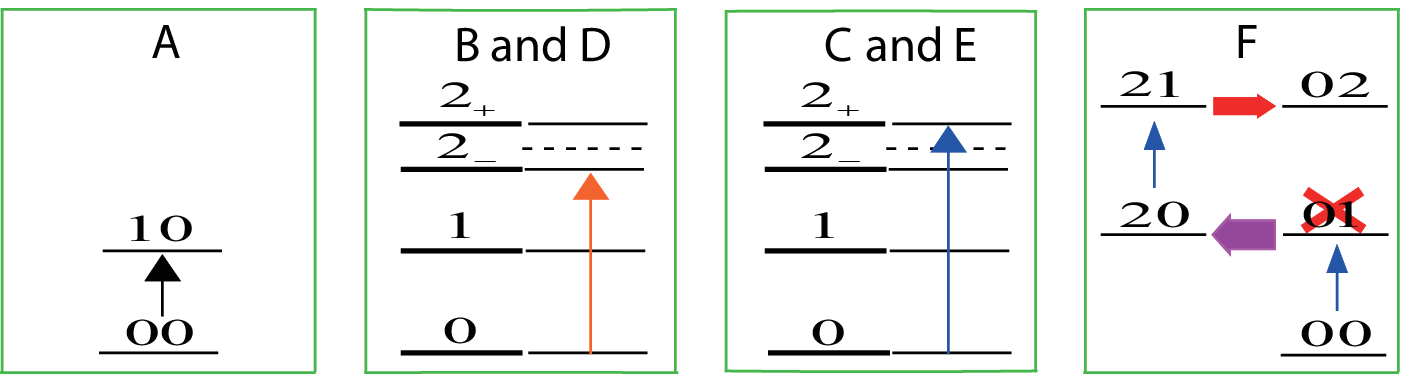}}
\caption{(color online) \textbf{Rescaled  average photon} number
(black ``$\Box$") and the equal-time second-order correlation
function (blue ``$\circ$") as a function of detunings. {\it The
photon number is rescaled by $F^2$ to fit the correlation function}.
(a) and (b) are for the case with cavity $a$ pumped, and (c) and (d)
for cavity $b$ pumped. Once one cavity is pumped, another is left
free. Dotted lines are analytical results [see
Eqs.~(\ref{nans})-(\ref{gagbleft})]. Red-solid lines show the
numerical  results. In all plots, we chose $\Omega /\kappa =
4,F/\kappa = 0.2,{\omega _a}/\kappa  = 1,{\omega _b}/\kappa  =
2,{\kappa _a}/\kappa  = 1$. A, B, C, D, E, and F mark the maximum
and minimum, the corresponding  detunings are,   A: ${\Delta
_a/\Omega} = 0$, B: $ - \frac{1}{{\sqrt 2 }}$, C: $\frac{1}{{\sqrt 2
}}$, D: $\Delta_b/\Omega=-\sqrt 2 $, E: $\sqrt 2 $, and F: $0$. The
tiny deviation of the analytical results from the numerical ones
caused by the approximation $F/\kappa \ll 1$ used in
Eqs.~(\ref{nans})-(\ref{gagbleft}). A-F in the bottom panels
illustrate the transitions that may lead to the features seen at the
points A-F.  Suppression of the steady-state population of the level
$|01\rangle$  is indicated by a red $X$. More understanding of the
features can be found in the main text.}\label{blockade:}
\end{figure}

\begin{equation}
\begin{aligned}
g_j^{(2)}(0) = \frac{{\left\langle {\hat h{{_j^\dag }^2} \hat h_j^2}
\right\rangle }}{{{{\left\langle {\hat h_j^\dag {{\hat h}_j}}
\right\rangle }^2}}},\label{gj}
\end{aligned}
\end{equation}
where all operators are evaluated at the same instance of  time. In
Fig.~\ref{blockade:} we show  $g_a^{(2)}(0)$ and $g_b^{(2)}(0)$ as a
function of  detuning.  The figures are plotted  in the weak driving
limit. Interesting features can been found at detunings $0, -
\frac{1}{{\sqrt 2 }} ,\frac{1}{{\sqrt 2 }}, - \sqrt 2$, $ \sqrt 2$,
and $0$  marked respectively by   $A, B$, $C$, $D$, $E$, and $F$.
Recall that $g_j^{(2)}(0) < 1$ indicates photon antibunching, and
$g_j^{(2)}(0) \to 0$ implies complete photon blockade, we find that
the  point $A$ exactly corresponds to the  single-photon blockade,
similar to that in Kerr-type \cite{Werner610118011999,
Verger731933062006,Ferretti850333032012} or QED
\cite{Tian46R68011992,Werner610118011999,
Brecha5923921999,Rebic14901999,Rebic650638042002,
Kim3975001999,Smolyaninov881874022002} nonlinear systems.
Remarkably,  when the cavity $b$ is pumped, two-photon blockade for
cavity $a$ occurs. This can be found at points $D$ and $E$ in
Fig.~\ref{blockade:} (c)]. Of cause for the cavity $b$, it is still
a single photon blockade. We refer to this effect as the two-photon
blockade from the aspect of cavity $a$, which means that the
two-photon Fock states blockade the generation of more photons.

To gain more  insights into the single- and two-photon  blockade
shown in Fig.~\ref{blockade:}, we develop an approximately analytic
expression  for the system by considering only the eight lowest
energy levels  in Fig.~\ref{setupsssss:} (b). Assume  that the
system is initially prepared in $\left| {00} \right\rangle$, and
only these levels  are  occupied due to the pumping of  the driving,
the state of the system can be written as,
 \cite{Carmichael82731991}
\begin{equation}
\begin{aligned}
\left| {\Phi } \right\rangle  =& {C_{00}}\left| {00}
 \right\rangle  + {C_{10}}\left| {10} \right\rangle
 + {C_{01}}\left| {01} \right\rangle
 + {C_{20}}\left| {20} \right\rangle \\
&+{C_{11}}\left| {11} \right\rangle
+ {C_{30}}\left| {30} \right\rangle
+ {C_{21}}\left| {21} \right\rangle  + {C_{02}}\left| {02} \right\rangle,
\label{phit}
\end{aligned}
\end{equation}
besides, we take an effective Hamiltonian,  ${{\hat H}_{eff}} =
{{\hat H}_S} - i[{\kappa _a}{{\hat a}^\dag }\hat a + \kappa {{\hat
b}^\dag }\hat b]/2$ to describe the system.   This approach allows
us to evaluate the mean occupation numbers up to the order of
${F^2}$ and the second-order correlation function up to the order of
${F^4}$. By substituting  $\left| {\Phi } \right\rangle$ into the
Schr\"{o}dinger equation $i{\partial _t}\left| \Phi \right\rangle =
{{\hat H}_{eff}}\left| \Phi \right\rangle $ ($\hbar = 1$,
hereafter), we find the following coupled equations,
\begin{equation}
\begin{aligned}
{{\dot C}_{00}} =& 0,\\
{{\dot C}_{10}} =&  - iF{C_{11}} - \delta_a{C_{10}},\\
{{\dot C}_{20}} =&  - i\sqrt 2 \Omega {C_{01}} - 2\delta_a{C_{20}},\\
{{\dot C}_{01}} =&  - i\sqrt 2 \Omega {C_{20}} - iF{C_{00}} - \delta_b{C_{01}},\\
{{\dot C}_{11}} =&  - i\sqrt 6 \Omega {C_{30}} - iF{C_{10}} - (\delta_a + \delta_ b){C_{11}},\\
{{\dot C}_{30}} =&  - i\sqrt 6 \Omega {C_{11}} - 3\delta_a{C_{30}},\\
{{\dot C}_{21}} =&  - i2\Omega {C_{02}} - iF{C_{20}} - (2\delta_a + \delta_b){C_{21}},\\
{{\dot C}_{02}} =&  - i2\Omega {C_{21}} - i\sqrt 2 F{C_{01}} - 2\delta_b{C_{02}},
\label{rightp}
\end{aligned}
\end{equation}
where $\delta_j =\kappa_j/2+{\Delta _j}i $,  and the cavity $b$
being pumped was assumed. At steady states, these amplitudes
$C_{ij}$ do not evolve, the mean occupation numbers in this case are
${{\bar n}_a} = 2{\left| {{{\bar C}_{20}}} \right|^2}$ and ${{\bar
n}_b} = {\left| {{{\bar C}_{01}}} \right|^2}$, where ${\bar C}_{ij}$
denote the amplitude at steady-state. Up to first order in
$F/\kappa$ (see Appendix A), the mean photon number is,
\begin{equation}
\begin{aligned}
\frac{{{{\bar n}_a}}}{{{n_0}}} =& \frac{{2{\Omega ^2}}}{{S\left( {\frac{1}{4}, - {\Delta _b},1,\frac{{{\Delta _b}}}{2},\frac{1}{2}} \right)}},\\
\frac{{{{\bar n}_b}}}{{{n_0}}} =& \frac{{4S\left( {1,0,0,0,2} \right)}}{{S\left( {1, - 4{\Delta _b},4,2{\Delta _b},2} \right)}},
\label{nans}
\end{aligned}
\end{equation}
where $S\left( {a,b,c,d,e} \right) = {(a{\kappa _a} + b{\Delta _a} +
c{\Omega ^2})^2} + {(d{\kappa _a} + e{\Delta _a})^2}$ and $ {n_0} =
{(F/\kappa )^2}$. From   $S\left( {a,b,c,d,e} \right)$  in these
equations, we can obtain the location for local maxima and minima of
the average photon numbers, which are in excellent agreement with
the numerical results shown in Fig.~\ref{blockade:}. The eight-level
model also provides us with the analytical expression of the
second-order correlation functions (see Appendix A),
\begin{equation}
\begin{aligned}
g_a^{(2)}(0) =& \frac{{S\left( {\frac{1}{4}, - {\Delta _b},1,\frac{{{\Delta _b}}}{2},\frac{1}{2}} \right)}}{{4{F^2}{\Omega ^2}}},\\
g_b^{(2)}(0) =& \frac{{S\left( { - \frac{1}{2}{a_0},2{\Delta _a} + {\Delta _b},1,2{\Delta _a} + \frac{1}{2}{\Delta _b},\frac{1}{2}} \right)}}{{S\left( { - \frac{1}{2},2{\Delta _b}, - 2 + \frac{{ - 1 + 4\Delta _b^2}}{{4{\Omega ^2}}},{\Delta _b},1 + \frac{{{\Delta _b}}}{{{\Delta _a}}}} \right)}}\\
&\cdot\frac{{S\left( {1, - 4{\Delta _b},4,2{\Delta _b},2} \right)}}{{{S^2}\left( {1,0,0,0,2} \right)}},
\label{gagb}
\end{aligned}
\end{equation}
where ${a_0} = {\kappa _a} + 1/2$. By the same approach, we  can
obtain the  following   equations  for the case of   cavity $a$
being pumped,
\begin{equation}
\begin{aligned}
{{\dot C}_{00}} =& 0,\\
{{\dot C}_{10}} =&  - iF{C_{00}} - i\sqrt 2 F{C_{20}} - {\delta _a}{C_{10}},\\
{{\dot C}_{20}} =&  - i\sqrt 2 \Omega {C_{01}} - 2{\delta _a}{C_{20}} - i\sqrt 2 F{C_{10}},\\
{{\dot C}_{01}} =&  - i\sqrt 2 \Omega {C_{20}} - {\delta _b}{C_{01}},\\
{{\dot C}_{11}} =&  - i\sqrt 6 \Omega {C_{30}} - iF{C_{01}} - i\sqrt 2 F{C_{21}} \\
&-({\delta _a} + {\delta _b}){C_{11}},\\
{{\dot C}_{30}} =&  - i\sqrt 6 \Omega {C_{11}} - 3{\delta _a}{C_{30}} - i\sqrt 3 F{C_{20}},\\
{{\dot C}_{21}} =&  - i2\Omega {C_{02}} - i\sqrt 2 F{C_{11}}
- (2{\delta _a} + {\delta _b}){C_{21}},\\
{{\dot C}_{02}} =&  - i2\Omega {C_{21}} - 2{\delta _b}{C_{02}}.
\label{eqleft}
\end{aligned}
\end{equation}
The steady-state amplitudes (see Appendix A) are given by,
\begin{equation}
\begin{aligned}
\frac{{{{\bar n}_a}}}{{{n_0}}} =& \frac{4}{{S\left( {1,0,0,0,2} \right)}},\\
\frac{{{{\bar n}_b}}}{{{n_0}}} =& \frac{{{{\left( {F\Omega } \right)}^2}}}{{S\left( {\frac{1}{2},0,0,0,1} \right) \cdot S\left( {\frac{1}{4}, - {\Delta _b},1,\frac{{{\Delta _b}}}{2},\frac{1}{2}} \right)}},\\
g_a^{(2)}(0) =& \frac{{S\left( {1,0,0,0,2} \right) \cdot S\left( {\frac{1}{{{\kappa _a}}},0,0,\frac{{2{\Delta _b}}}{{{\kappa _a}}},0} \right)}}{{S\left( {1, - 4{\Delta _b},4,2{\Delta _b},2} \right)}},\\
g_b^{(2)}(0) =& \frac{{S\left( {1,0,0,0,2} \right) \cdot S\left( {\frac{1}{4}, - {\Delta _b},1,\frac{{{\Delta _b}}}{2},\frac{1}{2}} \right)}}{{S\left( {{a_1},{b_1},{c_1},{d_1},{e_1}} \right) \cdot {S^{ - 1}}\left( {\frac{{{\Delta _b}}}{{{\kappa _a}}},1,0,\frac{1}{2},\frac{1}{{2{\Delta _a}}}} \right)}},
\label{gagbleft}
\end{aligned}
\end{equation}
where,
\begin{equation}
\begin{aligned}
{a_1} =&  - \frac{1}{{16}} - \frac{{3{\kappa _a}}}{{16}}
- \frac{{\kappa _a^2}}{8} + \frac{{3\Delta _b^2}}{4} + \frac{{3{\kappa _a}\Delta _b^2}}{4},\\
{b_1} =& \frac{{3{\Delta _a}}}{4} + \frac{{3{\kappa _a}{\Delta _a}}}{2}
- 3{\Delta _a}\Delta _b^2 - \Delta _b^3 + {\Delta _b}{f_1},\\
{c_1} =&  - \frac{1}{2} - \frac{{3{\kappa _a}}}{2} - \frac{{\kappa _a^2}}{2}
+ 2\Delta _a^2 + 2\Delta _b^2 - 4{\Omega ^2},\\
{d_1} = & - \frac{{3{\Delta _a}}}{4} - 3{\Delta _b}{\Omega ^2}
- \frac{3}{4}{\kappa _a}{\Delta _a} + \frac{{\Delta _b^3}}{2} + {\Delta _b}{g_1},\\
{e_1} =&  - \frac{1}{8} + \Delta _a^2 + 3{\Delta _a}{\Delta _b}  +
\frac{3}{2}\Delta _b^2 + {\Omega ^2}{h_1}. \label{coefff}
\end{aligned}
\end{equation}
Here, ${f_1} = \frac{3}{4} + 3{\kappa _a} + \frac{{3\kappa _a^2}}{2}
- 2\Delta _a^2 + 6{\Omega ^2},{g_1} =  - \frac{3}{8} -
\frac{3}{4}{\kappa _a} - \frac{1}{4}\kappa _a^2 + 3\Delta _a^2 +
3{\Delta _a}{\Delta _b},{h_1} =  - 3 - 2\frac{{{\Delta
_b}}}{{{\Delta _a}}} - 2{\kappa _a}.$ Similarly,  by analyzing $S(a,
b, c, d, e)$ in Eqs.~(\ref{nans})-(\ref{gagbleft}), we can obtain
interesting points in $g_{a,b}^{(2)}(0)$ labeled by A, B, C, D, E,
and F, which are in excellent agreement with the numerical results
shown in Fig. ~\ref{blockade:}.

We now discuss the features observed in Fig.~\ref{blockade:}. We
will use the  eight-level model together with the diagonal basis in
Eq.~(\ref{01}) to understand the physics behind the features.  In
(a) and (b), i.e., the pumping is  on microcavity $a$, when the
detuning is zero, ${\Delta _a} = 0$ (point A in
Fig.~\ref{blockade:}), we can see $g_a^{(2)}(0) \ll 1$, indicating
complete antibunching due to the suppressed two-photon process
(panel A in Fig.~\ref{blockade:}), this leads  to the single-photon
blockade. Physically, when the left cavity $a$ is pumped by a
incident laser of frequency $\omega_L$, a incident  photon  with the
the same frequency  as the cavity field $\omega_a$ will excite the
cavity from vacuum $\left| 0 \right\rangle $ to the first excited
state $\left| 1 \right\rangle $ [marked by the blue-dotted arrow in
Fig.~\ref{setupsssss:} (b)]. Once the cavity has a photon,  the
second photon at that frequency will be blocked because, due to the
nonlinearity of (\ref{H}), excitation of the system from $\left| 1
\right\rangle $ to $\left| {{2_ + }} \right\rangle $ or $\left| {{2_
- }} \right\rangle $ requires an additional energy $\hbar \sqrt 2
\Omega $, which cannot be supplied by the second photon. At the
detuning ${\Delta _a}/g = \pm \frac{1}{{\sqrt 2 }}$ (point B and C
of (a)), bunching happens due to the two-photon resonant transition
$\left| {0} \right\rangle \to \left| {{2_ + }} \right\rangle$ and$
\left| {{2_ - }} \right\rangle$ (panel B and C), the physics behind
this bunching is similar.

A similar story takes place when the microcavity $b$ is pumped, see
Fig.~\ref{blockade:} (c). At detuning ${\Delta _b}/g =  \pm \sqrt 2$
(point D and E of (c)), resonant transition $\left| {0}
\right\rangle \to \left| {{2_ + }} \right\rangle$ and $\left| {{2_ -
}} \right\rangle$ (panel D and E) occurs, indicating almost
completely antibunching for cavity mode $b$. Mathematically, we
find that both the single-photon occupation  amplitude ${\bar
C}_{10}$, the three-photon occupation amplitude ${\bar C}_{11}$ and
${\bar C}_{30}$ equals to zero, while the two-photon occupation
amplitude ${\bar C}_{01} $ and ${\bar C}_{20}$ are much larger than
that of four-photon under the weak driving limit $\left| {{\alpha
_j}{\alpha _k}} \right| \ll 1$, see Eq.~(\ref{Aamplitude}). The
point F in Fig.~\ref{blockade:} corresponds to detuning ${\Delta _b}
= 0$. $g_b^{(2)}(0) \gg 1$ at this point, indicating strong
bunching. This is due to destructive interference that suppresses
the population on $\left| {01} \right\rangle $ (panel F in
Fig.~\ref{blockade:}), steering the  system  into a dark state,
$\left| {dark} \right\rangle =  - \cos \varphi \left| {00}
\right\rangle  + \sin \varphi \left| {20} \right\rangle $, where
$\tan \varphi  = F/\Omega$. This is reminiscent of the
electromagnetically induced transparency
\cite{Weis33015202010,Lukin754572003}. Due to the weak coupling,
$\left| {20} \right\rangle $ is almost not populated when the system
occupies the dark state, this induces  the transition from
$|20\rangle$ to $\left| {21} \right\rangle $, which in turn is
strongly coupled to $\left| {02} \right\rangle $. The net result is
the transition with one photon is suppressed compared to the
two-photon transition, leading to the bunching at ${\Delta _b} = 0$.

To confirm this point, we now show that two-photon resonance is
absent at  ${\Delta_b} = 0$. At first glance, the level diagram in
Fig.~\ref{setupsssss:}(b) together with bunching at point F in
Fig.~\ref{blockade:} suggest a two-photon resonance at zero detuning
$\Delta_b = 0$, where the energy of the two-photon state $\left|
{02} \right\rangle $ is equal to the energy of two incident photons
in mode b. However, as discussed above, the strong bunching effect
at $\Delta_b = 0$ comes completely from the suppression of the
one-photon population. Further, we find that the expected two-photon
resonance is canceled by interference. This can be seen from a
second-order perturbative calculation of the two-photon Rabi
frequency $\omega _{00\rightarrow 02}^{(2)}$ for the transition
$\left| {00} \right\rangle \to \left| {02} \right\rangle $. The
two-photon state $\left| {02} \right\rangle $ can be populated by
the drive ${{\hat h}_{dr}} = F(\hat b + {{\hat b}^\dag })$ from
vacuum via two intermediate one-photon eigenstates $\left| {{2_ \pm
}} \right\rangle $ given by Eq.~(\ref{01}) with energies ${\omega
_{2 \pm }} = {\Delta _b} \pm \sqrt 2 \Omega $ in the rotating frame.
The resulting Rabi frequency is
\begin{eqnarray}
\omega _{00 \to 02}^{(2)} = \sum\limits_{m = {2_ + },{2_ - }}
{\left\langle {20} \right|{{\hat h}_{dr}}\left| m \right\rangle
\left\langle m \right|{{\hat h}_{dr}}\left| {00} \right\rangle
/{\omega _m}} ,\label{absence}
\end{eqnarray}
which vanishes at $\Delta {}_b = 0$ as a consequence of  destructive
interference between the two amplitudes. Although the exact
cancelation is lifted by including finite dissipation and the full
spectrum,  this simple argument shows that the expected two-photon
resonance at $\Delta {}_b = 0$ is strongly suppressed.

It is important to address that the photon blockade in the case of
cavity $b$ being driven is a two-photon blockade from the side of
cavity $a$, hence we   refer to this type of photon blockade as
two-photon blockade.

So far, we have demonstrated both analytically and numerically a
variety of quantum properties revealing the unique nature of the
single- and two-photon blockade. The results suggest that  by
manipulating the detuning, optical diodes may be realized in such a
system  based on the explicit single- and two-photon blockade.

\section{single-photon and two-photon diode}
By the  use of the blockade features of the photonic semiconductor
microcavities and the  $\chi^2$ nonlinearities, in this section we
present a scheme to  realize a single-photon and two-photon diode.
We define a rectifying factor as the normalized difference between
the two output currents corresponding to  the system being pumped
through the left and right cavities (indicated by the wave vectors
$k$ and $-k$, respectively) \cite{Mascarenhas130764932013}
\begin{eqnarray}
R = \frac{{2{N_b}(k) - {N_a}[ - k]}}{{2{N_b}(k) + {N_a}[ - k]}}.
\label{R}
\end{eqnarray}
Substituting Eqs.~(\ref{nans})-(\ref{gagbleft}) into Eq.~(\ref{R}),
we can obtain an analytical expression of the rectifying factor,
\begin{equation}
\begin{aligned}
R = \frac{{{\kappa F^2} - {\kappa
_a}S[\frac{1}{2},0,0,0,1]}}{{{\kappa F^2} + {\kappa
_a}S[\frac{1}{2},0,0,0,1]}}.\label{Ra}
\end{aligned}
\end{equation}
\begin{figure}[h]
\centering
\includegraphics[angle=0,width=0.495\textwidth]{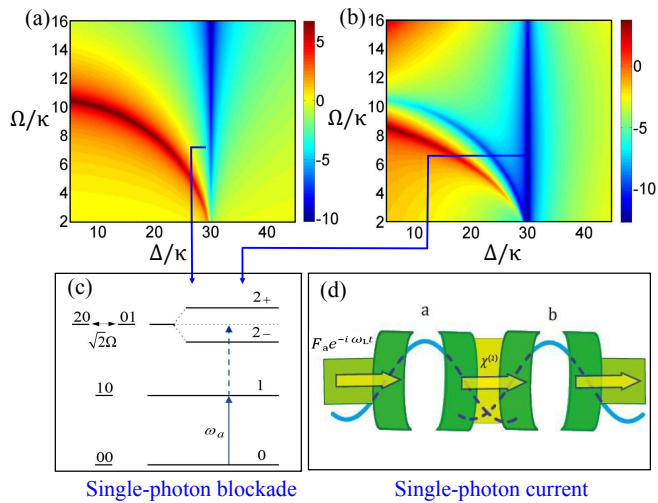}
\caption{(Color online) Single-photon rectification in the
semiconductor microcavities coupled via ${\chi ^2}$ nonlinearities.
The equal-time second-order correlation function  when the system is
pumped on the cavity $a$ [see analytical expression
Eq.~(\ref{gagbleft})] (log scale). (a) is  for the cavity $a$ and
(b) for the cavity $b$. We set $x = {\Delta _b} +2 {\Delta _a}$,
with detunings given by $\Delta  = {\Delta _b} - 2{\Delta _a}$
corresponding to ${\Delta _a} = (x - \Delta )/4$ and ${\Delta _b} =
(x + \Delta )/2$. Parameters chosen are ${\kappa _a} = 0.1\kappa
,{F} = 0.1\kappa,\Omega  = 10\kappa ,x = 30\kappa.$ The very narrow
region around  $\Delta  = 30\kappa $, i.e., ${\Delta _a} \approx 0$,
shows the complete single-photon blockade, the corresponding
transition is shown in   Fig.~\ref{leftjieshi:}-(c), and the
single-photon current to right cavity $b$ is illustrated in (d).}
\label{leftjieshi:}
\end{figure}
By this definition, $R=-1$ implies  maximal rectification with
enhanced transport to the left (left rectification), $R=0$ indicates
no rectification, while $R=1$ describes  maximal rectification with
transport to the right (right rectification).

In Fig.~\ref{leftjieshi:}, we plot the second-order correlation
function of the left cavity $a$ [see (a)] and right cavity $b$ [see
(b)] when the system is pumped on the cavity $a$. We find that the
normalized correlations of $g_j^{(2)}(0)  \ll 1$ reach their minimum
at $\Delta = 30\kappa$ (namely, ${\Delta _a} \approx 0$), implying
the well known single photon blockade. The diagram (c) in
Fig.~\ref{leftjieshi:} illustrates the transition  for this
blockade.

\begin{figure}[h]
\centering
\includegraphics[angle=0,width=0.49\textwidth]{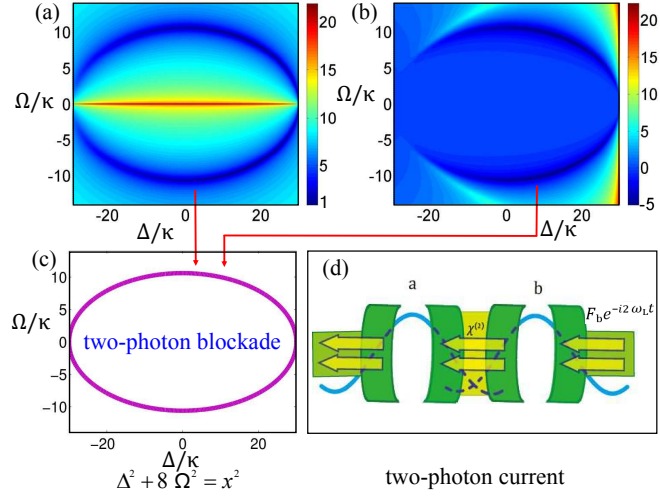}
\caption{(Color online) Two-photon rectification in the
semiconductor microcavities coupled by ${\chi ^2}$ nonlinearities.
The equal-time second-order correlation function[see analytical
expression Eq.~(\ref{gagb}), log scale]  of the cavity $a$ (see,
(a)) and of  the cavity $b$(see, (b)), this figure is plotted with
the right cavity $b$  pumped. We set $x = {\Delta _b} +2 {\Delta
_a}$ with detunings defined  by $\Delta  = {\Delta _b} -2 {\Delta
_a}$. Parameters chosen are ${\kappa _a} = 0.1\kappa ,{F} =
0.1\kappa,\Omega  = 10\kappa, x = 30\kappa.$ At the  ellipse
${\Delta ^2} + 8{\Omega ^2} = {x^2}$, i.e., two-photon resonant
transition from  $\left| 0 \right\rangle  \to \left| {{2_ \pm }}
\right\rangle $ when $2{\omega _L} = {\omega _{2 \pm }}$, two-photon
blockade happens  in the semiconductor microcavities, the
corresponding transition is illustrated  in
Fig.~\ref{rightjieshi:}-(c), this causes  the two-photon current to
the left cavity mode $a$, see (d). Note that the two-photon current
is symmetric for the symmetric coupling $\Omega$ in (a) and (b),
this is a direct reflection of  Eq.~(\ref{gagb}).}
\label{rightjieshi:}
\end{figure}

In Fig.~\ref{rightjieshi:}, we plot the correlation function of
cavity $a$ [see (a)] and cavity $b$ [see (b)] when the system is
pumped on the right cavity $b$. We observe that the minimum of
$g_b^{(2)}(0)$ is on an ellipse defined by Eq. (\ref{elliptic}) and
shown in Fig.~\ref{rightjieshi:} (c). This can be explained as the
condition for the  two-photon blockade to happen in the system,
i.e.,  two photon resonance $2{\omega _L}= {\omega _{2 \pm }}$
follows
\begin{eqnarray}
{\Delta ^2} + 8{\Omega ^2} = {x^2}, \label{elliptic}
\end{eqnarray}
where $x = {\Delta _b} + 2{\Delta _a}$ and  $\Delta ={\Delta _b} - 2
{\Delta _a}$. Analogously,  the bunching of a pair of photons for
the left cavity $a$ take place  due to the extreme antibunching of
the right cavity $b$.

\begin{figure}[h]
\centering
\includegraphics[angle=0,width=0.487\textwidth]{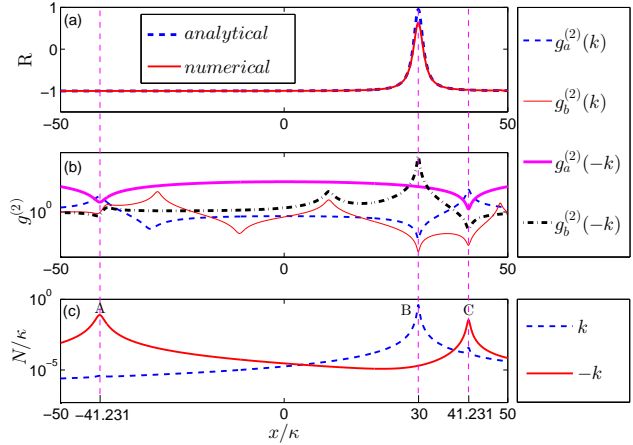}
\caption{(Color online) (a) The rectifying factor, correlation
function and  total current as a function of  frequency of the
external laser field $x$. The blue-dashed line and red-bold line
denote the rectification factor for the analytical expression
Eq.~(\ref{Ra}) and the fully numerical simulation in (a),
respectively. In (b), the equal-time second-order correlation
function $g^{(2)}$ [see analytical expression
Eqs.~(\ref{nans})-(\ref{gagbleft})] of the output photon in both
directions. (c) shows the total current, $N(k) = {N_a}(k) +
{N_b}(k)$ [see analytical expression
Eqs.~(\ref{nans})-(\ref{gagbleft})]. Parameters chosen are ${\kappa
_a} = 0.1\kappa, F = 0.1\kappa, \Omega = 10\kappa.$  We assume that
$ \Delta  = 30\kappa$ with ${\Delta _a} = (x - \Delta )/4$ and
${\Delta _b} = (x + \Delta )/2$.} \label{san:}
\end{figure}

Fig.~\ref{san:} shows $R$, $g^{(2)}$ and $N$ as a function of laser
frequency with a fixed $\Delta$ in the  small $\Omega/\kappa$ limit.
We find a local maximum of left rectification  in Fig.~\ref{san:}
(a) at $x=30\kappa$ (marked by $B$ in (c)), which corresponds to the
anti-bunching  in Fig.~\ref{san:} (b) at the left cavity $a$. The
total photon current $N(k) = {N_a}(k) + {N_b}(k)$  overcomes $N(-k)$
in the area around the $x = 30$, see  Fig.~\ref{san:} (c). This can
be explained as the single-photon blockade. From Eq.~(\ref{ll22}) we
can find the single-photon occupation  amplitude ${\bar C}_{10}$ is
sufficiently larger than that of multiphoton, e.g., two-photon
occupation  amplitude ${\bar C}_{01}$ and ${\bar C}_{20}$ in the
limit of weak optical driving, i.e., $\left| {{\alpha _j}{\alpha
_k}} \right| \ll 1,$.

Two maxima of left  rectification can be observed  at $x= \pm
10\sqrt {17}\kappa \approx  \pm 41.231\kappa$ (marked by $A$ and
$C$) from Fig.~\ref{san:} due to the two-photon blockade at the
ellipse Eq. (\ref{elliptic}), which corresponds to anti-bunched
effect in Fig.~\ref{san:} (b)  at the right cavity $b$ with  cavity
$b$ pumped (denoted by $-k$). The total photon current $N(-k)$
overcomes $N(k)$ in the area around the two points $A$ and $C$ in
Fig.~\ref{san:} (c). This observation can be explained by examining
Eq.~(\ref{Aamplitude}),  the contribution of the four-photon terms
${{\bar C}_{21}}$ and ${{\bar C}_{02}}$ can be neglected compared to
the two-photon occupation  amplitude ${{\bar C}_{20}}$ ${{\bar
C}_{01}}$ when the weak optical driving condition $\left| {{\alpha
_j} {\alpha _k}} \right| \ll 1,$ is satisfied.  The single-photon
occupation  amplitude ${{\bar C}_{10}}$ and three-photon occupation
amplitude ${{\bar C}_{11}}$ and ${{\bar C}_{30}}$ are equal to zero.
Therefore the single- and three-photon processes  have no
contribution  to the two-photon diode.

\section{delayed correlation function}
In addition to the equal-time second-order correlation functions
discussed above, quantum signatures can also be manifested in photon
intensity correlations with a finite time delay. This motivates  us
to investigate the dynamical evolution of the second-order
time-delayed  correlation function. The two-time intensity
correlations are defined by\cite{Liew1041836012010,
Verger731933062006,Ferretti820138412010},
\begin{equation}
\begin{aligned}
g_a^2(\tau ) = g_a^2(\tau  = {t_1} - t,t \to \infty )  =
\frac{{\left\langle {{{\hat a}^\dag }(t){{\hat a}^\dag }({t_1})\hat
a({t_1})\hat a(t)} \right\rangle }}{{{{\left\langle {{{\hat a}^\dag
}(t)\hat a(t)} \right\rangle }^2}}}. \label{delayedcorre}
\end{aligned}
\end{equation}
Rewriting  this correlation in terms of a classical  light photon
intensity $I$, $g^2(\tau ) = {{{\left\langle {I(\tau )I(0)}
\right\rangle } \mathord{\left/
 {\vphantom {{\left\langle {I(\tau )I(0)} \right\rangle }
 {\left\langle I \right\rangle }}} \right.
 \kern-\nulldelimiterspace} {\left\langle I \right\rangle }}^2}$
and using the Schwarz inequality, we obtain the
inequality \cite{Brecha5923921999,Carmichael82731991}
\begin{equation}
\begin{aligned}
{g^2}(\tau ) \le& {g^2}(0).
\label{sch}
\end{aligned}
\end{equation}
\begin{figure}[h]
\centering
\includegraphics[angle=0,width=0.52\textwidth]{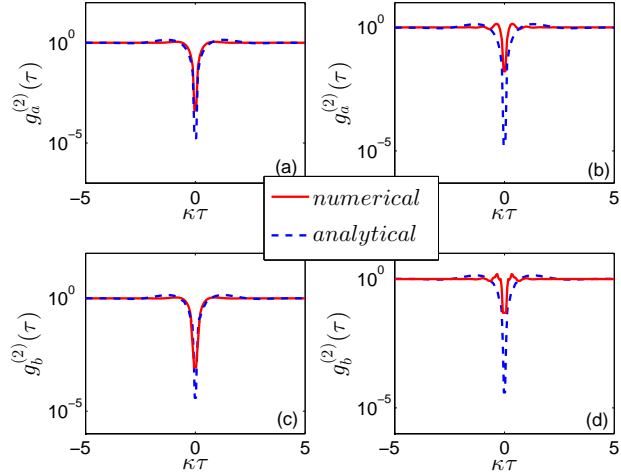}
\caption{(Color online) Finite temporal evolution of the
second-order correlation function for selected   driving strength
$F$ with cavity $a$ being pumped. The blue-dashed line and red-bold
line denote the analytical expression[given by Eq.~(\ref{B11}) and
(\ref{B22})]  and  the exact numerical simulation, respectively.
Parameters chosen are ${\kappa _a} = 4\kappa ,F = 1.8\kappa,\Omega =
25\kappa ,{\Delta _a} = 1.2\kappa,{\Delta _b} = 1\kappa $ for (a)
and (c), and $F=5\kappa$ for (b) and (d).} \label{delayed:}
\end{figure}
Similar to the  inequality ${g^{(2)}}(0) > 1$  at  zero time-delay,
violation of the inequality at finite delay is a signature of
quantum nature. We calculate the delayed second-order correlation
functions for both the left cavity $a$ and right cavity $b$ when the
system is pumped on the left cavity $a$.

The correlation function $g_a^{(2)}(\tau )$ and $g_b^{(2)}(\tau )$
are shown in Fig.~\ref{delayed:} for two strengthes  of the driving
$F$. We can understand the finite delay intensity correlations in
terms of the simple eight-level model discussed in the previous
section.  For this purpose, we rewrite Eq.~(\ref{delayedcorre}) as
\begin{equation}
\begin{aligned}
g_a^2({t_1} - t) = \frac{{T{r_S}T{r_E}[{{\hat a}^\dag }\hat
a{U^\dag }(t - {t_1})\hat a{\rho _{tot}}(t){{\hat a}^\dag }U(t -
{t_1})]}}{{{{\left\langle {{{\hat a}^\dag }(t)\hat a(t)}
\right\rangle }^2}}} \label{gatau}
\end{aligned}
\end{equation}
where the unitary evolution operator $U(t) =  \exp ( - i{\hat
H_T}t)$, ${\hat H_T} = {\hat H_S} + {\hat H_E} + {\hat H_I}$,  and
$\hat H_S$ is given by Eq.~(\ref{Hj}). $\hat H_E$ and $\hat H_I$ are
the Hamiltonians of the environment and the system-environment
interaction, respectively. We assume $\tau= {t_1}{\rm{ - }}t $ and $
t \to \infty,$ when the system density matrix arrive at a steady
state ${{\rho _s}}$. Applying  the Born approximation,    we have
\begin{equation}
\begin{aligned}
g_a^2(\tau ) = \frac{{Tr_STr_E[{\hat a^\dag }\hat a{U}(\tau )\hat a{\rho _s}
{\hat a^\dag } \otimes {\rho _E}U^\dag (\tau )]}}{{{{\bar
n}_a}^2}}.\label{steadytau}
\end{aligned}
\end{equation}
A  simple calculation follows,
\begin{eqnarray}
g_a^2(\tau ) = T{r_S}[{\hat A^\dag }\hat A{\rho
_{new}}(\tau)]\label{ga2tt},
\end{eqnarray}
where
\begin{eqnarray}
{\rho _{new}}(\tau) = T{r_E}[U(\tau ){\rho _{new}}(0) \otimes {\rho
_E}{U^\dag }(\tau )], \label{rho1t}
\end{eqnarray}
where the new initial state is defined by ${\rho _{new}}(0) =  \hat
a{\rho _s}{\hat a^\dag }$ and $\hat A = \hat a/{{\bar n}_a}.$ From
Eq.~(\ref{ga2tt}), we can find that the finite delayed second-order
correlation function can be thought of as an expectation values of
the effective photon number ${\hat A^\dag }\hat A$ with a new
density matrix given by Eq.~(\ref{rho1t}). Note that the new
dynamical  equation is the same as Eq.~(\ref{rho}), i.e., ${{\dot
\rho }_{new}}(t) \equiv \dot \rho (t)$, except that the new initial
condition $\rho (0)$ is replaced by ${\rho _{new}}(0)$.

Writing  the system steady state as ${\rho _s} =  \left| {\bar \Phi
} \right\rangle \langle \bar \Phi |$ and the new time-evolved
density matrix as ${\rho _{new}}(t) = \left| {{\Phi _{new}}(t)}
\right\rangle \left\langle {{\Phi _{new}}(t)} \right|$, we find
$\left| {{{\dot \Phi }_{new}}(t)} \right\rangle  \equiv \left| {\dot
\Phi (t)} \right\rangle $ satisfying Eq.~(\ref{eqleft}) within  the
 non-Hermitian Hamiltonian approximation. Therefore
the time-delayed correlation function can be calculated by
introducing   a new initial state
\begin{eqnarray}
\left| {{\Phi _{new}}(0)} \right\rangle  = \hat a\left| {\bar \Phi }
\right\rangle , \label{comditional}
\end{eqnarray}
with ${\left| {\bar \Phi } \right\rangle }$ given by
Eq.~(\ref{eqleft}) (see Appendix B). With this initial state, we
have the time-delayed correlation function,
\begin{equation}
\begin{aligned}
g_a^2(\tau ) = \frac{{{{\left| {{C_{10}}(\tau )} \right|}^2}}}
{{{{\left| {{{\bar C}_{10}}} \right|}^4}}} \label{B11}
\end{aligned}
\end{equation}
for the left cavity $a$, while for the right cavity $b$,
\begin{eqnarray}
g_b^2(\tau ) = \frac{{{{\left| {{C_{01}}(\tau )} \right|}^2}}}
{{{{\left| {{{\bar C}_{01}}} \right|}^4}}}, \label{B22}
\end{eqnarray}

We note the the non-Hermitian Hamiltonian  approximation is a good
approximation  when
\begin{eqnarray}
{\Omega ^2} \gg {\kappa _a} \kappa F.
\label{strong}
\end{eqnarray}

We find from Fig. \ref{delayed:}  that, when the strong coupling
conditions (\ref{strong}) are satisfied [see Figs.~\ref{delayed:}
(a) and (c)] and  the cavity $a$ is pumped, the finite delayed
second-order correlation functions given by the  analytical
expression Eqs.~(\ref{B11}) and (\ref{B22}) are in good agreement
with those   by   numerical simulations.   When the driving strength
is strong (\ref{strong}) [see Figs.~\ref{delayed:} (b) and (d)], the
analytical expression Eqs.~(\ref{B11}) and (\ref{B22})  deviates
from the  numerical simulations. In addition, from
Fig.~\ref{delayed:}, we can observe  that the $g_a^{(2)}(\tau )$ and
$g_b^{(2)}(\tau )$ increase above its initial value at finite delay.
This is an violation of the   inequality in Eq.~(\ref{sch}),
indicating that photons emitted at different times prefer to stay
together. See the red-bold line around $\tau \approx \pm
0.327/\kappa$, in    Fig.~\ref{delayed:}(b) and (d).

This tells  that two subsequent  emissions are suppressed in
single-photon diode, leading  to dynamical anti-bunching both for
zero- and finite-time delay. Hence our proposal provides us with new
insight on the necessary time delay in a single-photon diode in the
semiconductor microcavities coupled via $\chi^{(2)}$ nonlinearities.

\section{Conclusion}
In summary, we have presented a scheme for creating an optical
diode, the diode  is composed of two   microcavities coupled via
${\chi ^{(2)}}$ nonlinearities. A master equation  to describe such
a system is given. By Solving this master equation, single- and two
photon blockades are predicted. By the use of this photon blockade,
we design an optical diode, which has a merit to combine   photon
rectification and single photon to two-photon conversion. To
characterize the rectification, we calculated both analytically and
numerically the  two-order correlation function and the rectifying
factor in the weakly driven limit. The numerical simulation and
analytical expression agree well with each other.

\section*{ACKNOWLEDGMENTS}
We would like to thank Prof. W. T. M. Irvine for helpful
discussions. This work is supported by the NSF of China under Grant
No. 11175032.
\appendix
\section{Analytical  expression}
In this Appendix, we provide the analytical expression used to
calculate zero and finite delayed second-order correlation functions
in the steady state. First, zero delayed correlations are calculated
by the use of  steady-state solutions of Eq.~(\ref{rightp}). We set
the time derivatives to zero and solve the equations iteratively,
order by order, in the weak driving. With the right microcavity $b$
being pumped, simple calculation yields
\begin{equation}
\begin{aligned}
{{\bar C}_{00}} =& 1,{{\bar C}_{10}} = {C_{11}} = {C_{30}} = 0,\\
{{\bar C}_{20}} =& \frac{{{\alpha _a}{x_b}}}{{\sqrt 2 (1 - {x_a}{x_b})}},\\
{{\bar C}_{01}} =&  - \frac{{{\alpha _b}}}{{1 - {x_a}{x_b}}},\\
{{\bar C}_{21}} =& \frac{{ - {\alpha _a}{\alpha _b}(2 + y){x_b}}}
{{\sqrt 2 (2 + y - 2{x_a}{x_b})(1 - {x_a}{x_b})}},\\
{{\bar C}_{02}} =& \frac{{{\alpha _a}{\alpha _b}(2 + y + {x_a}{x_b}y)}}
{{\sqrt 2 y(2 + y - 2{x_a}{x_b})(1 - 2{x_a}{x_b})}},
\label{Aamplitude}
\end{aligned}
\end{equation}
where ${\alpha _j} = -iF/{\delta _j} ({\left| {{\alpha _j}}
\right|^2}$ or $\left| {{\alpha _j}{\alpha _k}} \right| \ll 1),{x_j}
= -i\Omega /{\delta _j}$, and $y = {\delta _b}/{\delta _a}$. Using
these amplitudes, we can express all equal-time averages. The mean
photon number is,
\begin{equation}
\begin{aligned}
{{\bar n}_a} =& 2{\left| {{{\bar C}_{20}}} \right|^2},\\
{{\bar n}_b} =&{\left| {{{\bar C}_{01}}} \right|^2},
\label{bana}
\end{aligned}
\end{equation}
and the photon-photon correlation functions are
\begin{equation}
\begin{aligned}
g_b^{(2)}(0) = \frac{{2{{\left| {{{\bar C}_{02}}} \right|}^2}}}
{{{{\left| {{{\bar C}_{01}}} \right|}^4}}},\\
g_a^{(2)}(0) = \frac{1}{{2{{\left| {{{\bar C}_{20}}} \right|}^2}}}.
\label{banagana}
\end{aligned}
\end{equation}
Therefore Eqs.~(\ref{nans}) and (\ref{gagb}) can be obtained by
substituting Eq.~(\ref{Aamplitude}) into Eqs.~(\ref{bana}) and
(\ref{banagana}), respectively. Next we calculate these occupation
amplitudes for pumping is on the left cavity $a$,
\begin{eqnarray}
{{\bar C}_{00}} &=& 1,{{\bar C}_{10}} =  - {\alpha _a},\nonumber\\
{{\bar C}_{01}} &=& \frac{{{\alpha _a}{\alpha _b}{x_a}}}{{({x_a}{x_b} - 1)}},\nonumber\\
{{\bar C}_{20}} &=&  - \frac{{\alpha _a^2}}{{\sqrt 2 ({x_a}{x_b} - 1)}},\nonumber\\
{{\bar C}_{30}} &=& \frac{{\alpha _a^2{\alpha _b}(2 + 3y + {y^2} + 2{x_a}{x_b} - 4x_a^2x_b^2)}}{{\sqrt 6 \lambda ({x_a}{x_b} - 1)}},\nonumber\\
{{\bar C}_{11}} &=& \frac{{ - {\alpha _a}\alpha _b^2{x_a}(2 + 3y + {y^2} - 2{x_a}{x_b} - 2x_a^2)}}{{\lambda ({x_a}{x_b} - 1)}},\nonumber\\
{{\bar C}_{21}} &=& \frac{{\sqrt 2 \alpha _a^2\alpha _b^2{x_a}(1 + y)}}{{\lambda ({x_a}{x_b} - 1)}},\nonumber\\
{{\bar C}_{02}} &=& \frac{{ - \sqrt 2 \alpha _a^2\alpha _b^2{x_a}{x_b}(1 + y)}}{{\lambda ({x_a}{x_b} - 1)}},
\label{ll22}
\end{eqnarray}
where the coefficient $\lambda  = (2{y^{ - 1}}  + 3 + y - 2x_b^2 -
6{x_a}{x_b} - 2x_a^2 + 4x_a^2x_b^2).$ The mean photon numbers  and
the photon-photon correlation functions are
\begin{equation}
\begin{aligned}
{{\bar n}_b} =& {\left| {{{\bar C}_{01}}} \right|^2},\\
{{\bar n}_a} =& {\left| {{{\bar C}_{10}}} \right|^2},\\
g_b^{(2)}(0) =& \frac{{2{{\left| {{{\bar C}_{02}}} \right|}^2}}}
{{{{\left| {{{\bar C}_{01}}} \right|}^4}}},\\
g_a^{(2)}(0) =& \frac{{2{{\left| {{{\bar C}_{20}}} \right|}^2}}}
{{{{\left| {{{\bar C}_{10}}} \right|}^4}}}.
\label{gfgnb}
\end{aligned}
\end{equation}
Eq.~(\ref{gagbleft}) can be obtained by  substituting
Eq.~(\ref{ll22}) into Eqs.~(\ref{gfgnb}).

\section{delayed correlation function}

The dynamical evolution of the second-order correlation function can
be calculated within  the same approach as in Appendix A. By the use
of Eq.~(\ref{comditional}) as the initial condition and consider a
pumping is on the left cavity $a$. Specifically, the unnormalized
state after the annihilation of a photon in the left cavity $a$ is
$a\left| {\bar \Phi } \right\rangle = {{\bar C}_{10}}\left| {00}
\right\rangle  + \sqrt 2 {{\bar C}_{20}}\left| {10} \right\rangle$,
where we have ignored the high-order terms. With this understanding
in mind, Eq.~(\ref{B11}) can be obtained by solving the first four
equations of Eq.~(\ref{eqleft}) for the amplitudes with  the initial
condition.

The correlation function of the right cavity $b$ mode $g_b^2(\tau )$
may be calculated similarly.  The state after annihilation of a
photon in the $b$ mode is $b\left| {\bar \Phi } \right\rangle =
{{\bar C}_{01}}\left| {00} \right\rangle  + {{\bar C}_{11}}\left|
{10} \right\rangle  + {{\bar C}_{21}}\left| {20} \right\rangle  +
\sqrt 2 {{\bar C}_{02}}\left| {01} \right\rangle.$ Using this as the
initial condition, Eq.~(\ref{B22}) can be obtained by solving
Eq.~(\ref{eqleft}).

\end{document}